\documentclass[copyright,creativecommons]{eptcs}
\providecommand{\event}{LSFA 2016}
\usepackage[T1]{fontenc}
\usepackage[utf8]{inputenc}
\usepackage{graphicx} 
\usepackage{amsmath}
\usepackage{amsthm}
\usepackage{amssymb}
\usepackage{semantic}
\usepackage{mathpartir}
\usepackage{hyperref}
\usepackage{appendix}
\usepackage{verbatim} 
\usepackage[dvipsnames]{xcolor}
\usepackage{tikz}

\newtheorem{definition}{Definition}
\theoremstyle{definition}
\newtheorem{example}{Example}

\title{MaudeTypedLog: A Typed Interpreter for Prolog in Maude \thanks{This work has been partially supported by the grant PID2024-162030OB-100 funded by CIN/AEI/10.13039/501100011033 and ERDF A way of making Europe, by the grant CIPROM/2022/6 funded by Generalitat Valenciana, and by the NATO Science for Peace and Security Programme project SymSafe (grant number G6133).}}
\author{Enrique Gallifa-Tronch
\institute{VRAIN, Universitat Politècnica de València, Camí de Vera, s/n, 46022}
\email{egaltro@upv.es}
\and
João Barbosa
\institute{DCC, Faculdade de Ciências da Universidade do Porto, Rua do Campo Alegre s/n, 4169-007}
\institute{LIACC - Artificial Intelligence and Computer Science Laboratory}
\email{joao.barbosa@fc.up.pt}
\and
Santiago Escobar
\institute{VRAIN, Universitat Politècnica de València, Camí de Vera, s/n, 46022}
\email{sescobar@upv.es}}
\def\titlerunning{Typed Interpreter in Maude}
\def\authorrunning{Gallifa-Tronch, Barbosa, Escobar}

\begin{document}

\maketitle

\begin{abstract}
Prolog is traditionally thought of as an untyped logic programming language, although there are queries that result in a type error. Several attempts of statically introducing a type discipline in Prolog have been made but they have not been widely adopted.
We use Maude to implement a typed unification algorithm and use it as the basis for an interpreter for Prolog called MaudeTypedLog. This interpreter follows the Typed SLD-resolution operational semantics for logic programming, that makes it possible to detect type errors in both programs and queries dynamically.

\end{abstract}

\section{Introduction}

Prolog is considered an untyped programming language, since there is no type information when defining programs or executing them, in \textit{pure} Prolog. However, in \textit{full} Prolog, several built-in predicates have some type information such as \texttt{int/1}, \texttt{list/1}, and \texttt{float/1} which check that terms have specific types, and \texttt{=../2} that expects the right-hand side argument to be a list. Therefore, for certain queries, we can get a \textit{type error} as the output.

Since there is already some type consideration in Prolog compilers, one could try to extend this to other definitions. Typical examples are \texttt{append/3} and \texttt{member/2}: one could say that the query \texttt{member(1,2)} returns \texttt{no}, as does \texttt{member(2,[1,3])}, but for different reasons -- one is a type error, the other is saying that \texttt{2} does not occur in the list \texttt{[1,3]}; and the fact that \texttt{append([],1,1)} returns \texttt{true} goes against any specification of \texttt{append/3}, which is supposed to represent the concatenation of two lists into a third. 


Several authors \cite{DBLP:conf/iclp/Zobel87,Mishra84,DBLP:books/mit/pfenning92/YardeniFS92,Fruhwirth88,DBLP:books/mit/pfenning92/DartZ92,schrijvers2008towards,gallagher2004abstract,BarbosaFloridoCosta21,Hermenegildo00C23} have defined a type discipline for Prolog. Most of the type languages are based on regular types, which are types described by regular term grammars \cite{DBLP:conf/iclp/Zobel87,Mishra84,DBLP:books/mit/pfenning92/YardeniFS92,Fruhwirth88,DBLP:books/mit/pfenning92/DartZ92,BarbosaFloridoCosta21}. A subset of regular types, called \textit{deterministic} regular types, allows for decidable emptiness checking, subset checking, and intersection and unification operations as defined by Dart and Zobel \cite{DBLP:books/mit/pfenning92/DartZ92}.

Types can be used as verification tools and some approaches to types in logic programming try to detect type errors at compile time using assertions \cite{DBLP:journals/tplp/HermenegildoBCLMMP12}, type inference \cite{DBLP:conf/slp/PyoR89,BarbosaFloridoCosta21,BarbutiG92,DBLP:conf/iclp/Zobel87,Fruhwirth88,BruynoogheGallagherWellTyping,Vaucheret:2002:MPY:647171.718317}, or type declarations \cite{DBLP:journals/ai/MycroftO84,DBLP:conf/slp/LakshmanR91}. However, the approaches where types are declared require extra definitions by the authors, since they demand type declarations, or type assertions, for each predicate in the program. On the other hand, the type inference approaches often result in very general and vague types that allow for the detection of few type errors \cite{BarbosaFloridoCosta21}. In particular, in the type assertion language of Ciao-Prolog~\cite{DBLP:journals/tplp/HermenegildoBCLMMP12}, the assertions about types can be very specific, but they will be verified as assertions, so if they fail, we do not have information about which clause of the program causes the ``error". On the other hand, SICStus Prolog~\cite{CarlssonM12} has a library \texttt{types} that provides type checking, where the types can be user-defined using a regular type language, similar to \cite{barbosa2024iclp}. However, a lot of checks are made that we argue are orthogonal to type checking, for instance, we can check for uninstantiated variables, and such a variable having type \texttt{int} would be considered a type error. We argue this is not really type checking as we define it, and those are not the type errors we are concerned about.

One way to reduce the amount of declarations provided by the programmer would be to incentivize the declaration not of the type for every predicate, but of the algebraic data types that occur in a program, or collection of programs, similar to \texttt{data} declarations in Haskell \cite{Haskell}. This is described in \cite{barbosa2024iclp}. In that paper, the authors also describe a typed unification algorithm that detects type errors during unification. The traditional operational semantics for Prolog is SLD-resolution \cite{Apt:1996:LPP:249573}, where an atom of the query is selected, together with the head of a clause, for unification. Unification is, thus, the basic computing operation. The typed unification algorithm described in \cite{barbosa2024iclp} could be used as the unification algorithm in a typed operational semantics as the one described in \cite{BarbosaFloridoCosta22}.

Maude~\cite{CDELMMT:2007-book,DEEMMRTJLAMP20} is a high-performance logical framework 
based on rewriting logic \cite{unified-tcs}. 
Maude is routinely used for
formal specification \cite{MeseguerFASE25}, 
verification \cite{MeseguerLOPSTR25},
and declarative programming \cite{DEEMMRTPPDP24}.
Some examples are Petri nets, process calculus, object-based systems, asynchronous hardware, mobile ad hoc network protocol, cloud-based storage systems, web browsers, programming languages with threads, distributed  control systems, and models of mammalian cell pathways 
(see \cite{DEEMMRTJLAMP20} and references therein).
Maude is specially suited for 
(i) bridging the gap between formal specifications and code implementations; 
(ii) supporting formal verification and analysis from inductive invariants to infinite-state model checking;
(iii) early integration of Formal Methods in software design and deployment;
(iv) declarative programming of systems with heterogeneous components thanks to the use of external objects and reflection, 
e.g. a socket object, an I/O object, or a meta-interpreter object. 

Despite all these applications, Maude has not extensively been used for logic programming~\cite{EscobarALP23}.
Several versions of a Prolog interpreter were provided
in \cite{DEEMMRTJLAMP20}
using transition rules, equational rewriting and axioms such as associativity and commutativity, including notions such as cut and negation as rewriting execution strategies.
In this paper, we 
present \textsf{MaudeTypedLog}
that
(i) implements a typed three-valued unification algorithm 
\cite{barbosa2024iclp}
instead of an untyped one,
(ii) integrates the unification algorithm into one of the 
Prolog interpreters of \cite{DEEMMRTJLAMP20},
and
(iii) 
extends the SLD-resolution to provide
\textit{true}, \textit{false}, or \textit{wrong} results
according to the 
TSLD-resolution of \cite{BarbosaFloridoCosta22}. The types considered by the unification algorithm are parametrically polymorphic deterministic regular types.




\paragraph{Contributions:} The contributions of this paper are the following:
\begin{itemize}
    \item a first implementation of the typed unification algorithm described in \cite{barbosa2024iclp},
    \item a first implementation of an interpreter based on the TSLD-resolution operational semantics for Prolog, and
    \item a first implementation of an algorithm that uses TSLD-trees in order to detect type errors in programs and in queries. 
\end{itemize}

\section{TSLD-Resolution}

All of the definitions presented in this section come from previous papers by the authors \cite{BarbosaFloridoCosta22,barbosa2024iclp}, but we present them to make the paper self-contained. For a more detailed presentation and a semantic justification for these definitions, we refer the reader to the original papers.

\subsection{Terms and Types}

The language of terms that we will consider follows from Apt and Lloyd \cite{Apt:1996:LPP:249573,Lloyd:1984:FLP:2214}.
Given an infinite set of variables \textbf{Var} and an infinite set of function symbols \textbf{Func}, a term is:
\begin{enumerate}
    \item a variable in \textbf{Var} (\texttt{X}, \texttt{Y}, \texttt{Xi}, \dots);
    \item a function symbol in \textbf{Func} of arity 0 (\texttt{k}, \texttt{a}, \texttt{b}, \texttt{1}, \dots), which we call a constant;
    \item a function symbol in \textbf{Func} of arity $n\geq1$ (\texttt{f}, \texttt{g}, \texttt{h}, \dots) applied to an $n$-tuple of terms.
\end{enumerate}

Types are syntactic objects that are associated with semantic domains. We will present the necessary syntax for the types we will use throughout the paper. Assume an infinite set of type variables \textbf{TVar}, a finite set of base types \textbf{TBase}, an infinite set of typed function symbols \textbf{TFunc}, and an infinite set of type symbols \textbf{TSym}, parenthesis, and the comma. There is a one-to-one correspondence between \textbf{TFunc} and \textbf{Func} which we assume is predefined.
%
A type is usually represented as:
\begin{itemize}
\item a type variable in \textbf{TVar} ($\alpha$, $\beta$, $\alpha_i$, \dots),
\item a base type in \textbf{TBase} ($int$, $\mathit{float}$, $atom$, or $str$), or
\item a type symbol in \textbf{TSym} applied to an $n$-tuple of types ($list(int)$, \dots).
\end{itemize}
We assume every type symbol is defined by a deterministic type definition \cite{barbosa2024iclp}.

\subsection{Types of Terms}

Syntactic typing is defined by a type system. The type system defined in \cite{barbosa2024iclp} and presented again here is very simple. We assume a context $\Gamma$ -- a set of pairs $X : \tau$ that assign a type $\tau$ to each variable $X$ -- and a set of type assumptions for function symbols $\Delta$. Then we write $\Gamma,\Delta \vdash t:\tau$ to say that the term $t$ has type $\tau$, in context $\Gamma$ with assumptions $\Delta$. Assumptions in $\Delta$ are of the form $k:\forall \vec{\alpha}.\tau$, for constants, and $f : \forall \vec{\alpha}.\tau_1\times \dots \times \tau_n \to \tau$, for function symbols, where the generic variables $\vec{\alpha}$ of these type schemes are exactly the type variables that occur in $\tau$ and $\tau_1 \times \dots \times \tau_n \to \tau$, respectively. 
In order to be sure that $\Gamma,\Delta \vdash t:\tau$ we need to follow the rules of the type system (Figure \ref{TSystem}) and find a derivation for the statement.

\begin{figure*}[h]
\scalebox{.85}{
\begin{mathpar}

\inferrule* [left=\textsc{VAR}]
{(X:\tau) \in \Gamma }
{\Gamma, \Delta \vdash X:\tau}

\inferrule* [left=\textsc{CPL}]
{(f : \forall \vec{\alpha}.\tau_1 \times \dots \times \tau_n \to \tau) \in \Delta \\\\
\Gamma,\Delta \vdash t_1: \tau_1[\vec{\alpha} \mapsto \vec{\sigma}] ~~\dots ~~ \Gamma, \Delta \vdash t_n:\tau_n[\vec{\alpha} \mapsto \vec{\sigma}]}
{\Gamma, \Delta \vdash f(t_1,\dots,t_n) : \tau[\vec{\alpha} \mapsto \vec{\sigma}]}

\inferrule* [left=\textsc{CST}]
{(k : \forall \vec{\alpha}.\tau) \in \Delta}
{\Gamma, \Delta \vdash k:\tau[\vec{\alpha} \mapsto \vec{\sigma}] }

\inferrule* [left=\textsc{EQU}]
{\Gamma, \Delta \vdash t_1 : \tau ~~~~~ \Gamma, \Delta \vdash t_2 : \tau}
{\Gamma, \Delta \vdash t_1 = t_2 : bool}
\end{mathpar}}
\caption{Type System}
\label{TSystem}
\end{figure*}


\subsection{Constraints}
The typed unification algorithm will be defined as a rewrite system as in \cite{MartelliMontanari}. We will rewrite constraints over terms and over types.
The notation we use for equality constraints between terms is $t_1 = t_2$, and for equality constraints between types is $\tau_1 \doteq \tau_2$.

We say that a set of equality constraints is in {\em solved form} if all constraints are of the form $X_i = t_i$, for some term $t_i$, and there is no other occurrence of any $X_i$ anywhere else in the set. A set of equality constraints in solved form can be interpreted as a substitution, where every constraint of the form $X_i = t_i$ is interpreted as $[X_i \mapsto t_i]$. The concepts are similarly applicable to type equality constraints and we get type substitutions.

We assume that the reader is familiar with the common notions of substitutions and unifiers, such as composing two substitutions and applying substitutions to terms, or type substitutions to types.

The typed unification algorithm performs unification for terms and types. The intuition is that if the types do not unify, then there is a type error.
We follow the approach of \cite{Wand87}: generate constraints for typability and solve them. 

\subsection{Typed Unification Algorithm}

\begin{figure*}[ht]
\scalebox{.85}{
\begin{mathpar}
\inferrule* [left=\textsc{GVAR}]
{ (X: \alpha) \in \Gamma }
{\Gamma,\Delta \vdash X:\alpha ~|~ \emptyset~|~\emptyset}

\inferrule* [left=\textsc{GCST}]
{(k : \forall \vec{\alpha}.\tau) \in \Delta}
{\Gamma,\Delta \vdash k:\tau[\vec{\alpha} \mapsto \vec{\beta}]  ~|~ \emptyset ~|~ \emptyset}

\inferrule* [left=\textsc{GCPL}]
{(f : \forall \vec{\alpha}.\tau_1 \times \dots \times \tau_n \to \tau) \in \Delta \\ \\
\Gamma,\Delta \vdash t_1: \tau_1^{\prime} ~|~ \emptyset ~|~ T_1~~\dots ~~ \Gamma, \Delta \vdash t_n:\tau_n^{\prime}~|~ \emptyset ~|~ T_n}
{\Gamma, \Delta \vdash f(t_1,\dots,t_n) : \tau[\vec{\alpha} \mapsto \vec{\beta}]~|~\emptyset ~|~ T_1 \cup \dots \cup T_n  \cup \{\tau^{\prime}_1 \doteq \tau_1[\vec{\alpha} \mapsto \vec{\beta}], \dots, \tau^{\prime}_n \doteq \tau_n[\vec{\alpha} \mapsto \vec{\beta}] \}}

\inferrule* [left=\textsc{GEQU}]
{\Gamma,\Delta \vdash t_1 : \tau_1 ~|~C_1~|~T_1 ~~~~~ \Gamma,\Delta \vdash t_2 : \tau_2 ~|~C_2~|~T_2}
{\Gamma,\Delta \vdash t_1 = t_2 : bool ~|~C1 \cup C2 \cup \{t_1 = t_2\}~|~ T_1 \cup T_2 \cup \{\tau_1 \doteq \tau_2\}}
\end{mathpar}}
\caption{Constraint Typing Judgment -- $\beta$ are always fresh type variables}
\label{ConsGen}
\end{figure*}

The input to the typed unification algorithm, as with any unification algorithm, are two terms $t_1$ and $t_2$. However, the output may not be a unifier or false, instead we have a third value \textit{wrong} that will be the output in case the types do not unify. This is our way of detecting type errors -- they correspond to the unification of terms that cannot belong to the same semantics domain, which we detect by having non-unifiable types.

First, given the terms $t_1$ and $t_2$, we need to generate the sets of equality constraints and type equality constraints that will be rewritten by the algorithm. This generation is defined by the rules in Figure \ref{ConsGen}.

Given a derivation of $\Gamma,\Delta \vdash t_1=t_2|C|T$, we use the generated sets $C$ and $T$, of equality constraints and type equality constraints, respectively, and we apply the rules in the 
rewriting system 
of Figure~\ref{ConsSolve} in order.

\begin{figure}[ht]
\begin{enumerate}
    \item $(C, \{f(\tau_1,\dots,\tau_n) \doteq f(\tau^{\prime}_1,\dots,\tau^{\prime}_n)\} \cup \mathit{Rest}) \to (C, \{\tau_1 \doteq \tau^{\prime}_1,\dots,\tau_n \doteq \tau^{\prime}_n\} \cup \mathit{Rest})$
    \item $(C, \{\tau \doteq \tau\} \cup \mathit{Rest}) \to (C, \mathit{Rest})$
    \item $(C, \{f(\tau_1,\dots,\tau_n) \doteq g(\tau^{\prime}_1,\dots, \tau^{\prime}_m)\}\cup \mathit{Rest}) \to wrong$, if $f \neq g$ or $n \neq m$
    \item $(C, \{\tau\doteq\alpha\}\cup \mathit{Rest}) \to (C, \{\alpha\doteq\tau\}\cup \mathit{Rest})$, $\tau$ is not a type variable
    \item $(C, \{\alpha\doteq\tau\}\cup \mathit{Rest}) \to (C, \{\alpha\doteq\tau\}\cup \mathit{Rest}[\alpha \mapsto \tau])$, if $\alpha \notin \mathit{var}(\tau)$, $\alpha \in \mathit{var}(\mathit{Rest})$
    \item $(C, \{\alpha\doteq\tau\}\cup \mathit{Rest}) \to wrong$, if $\alpha$ occurs in $\tau$

    \item $(\{f(t_1,\dots,t_n) = f(s_1,\dots,s_n)\} \cup \mathit{Rest},T) \to (\{t_1 = s_1,\dots,t_n = s_n\} \cup \mathit{Rest},T)$
    \item $( \{t = t\} \cup \mathit{Rest},T) \to (Rest,T)$
    \item $(\{f(t_1,\dots,t_n) = g(s_1,\dots,s_m)\}\cup \mathit{Rest},T) \to false$, if $f\neq g$ or $n \neq m$
    \item $(\{t= X\}\cup \mathit{Rest},T) \to (\{X = t \}\cup \mathit{Rest},T)$, $t$ is not a variable
    \item $(\{X= t\}\cup \mathit{Rest},T) \to (\{X = t\}\cup \mathit{Rest}[X \mapsto t],T)$, if $X \notin  \mathit{var}(t)$, $X \in  \mathit{var}(\mathit{Rest})$
    \item $(\{X= t\}\cup \mathit{Rest},T) \to false$, if $X$ occurs in $t$.%
\end{enumerate}%
\caption{Typed Unification Algorithm rules}
\label{ConsSolve}
\end{figure}
Note that rules 1--6 correspond to the Martelli-Montanari algorithm \cite{MartelliMontanari} for type equality constraints, and rules 7--12 correspond to the same algorithm but for term equality. Also note that for the failure of unification of type equality constraints we get \textit{wrong} and for term equality we get \textit{false}, and since we apply the rules in order, we get \textit{wrong}, regardless of whether the terms would unify.

\subsection{Programs and Queries}

We now extend our language of terms to a language of programs by adding an infinite set of predicate symbols \textbf{Pred} and the reverse implication $\leftarrow$.

The definition of atoms, queries, clauses and programs is the usual one \cite{Apt:1996:LPP:249573}:
\begin{itemize}
    \item an atom is a predicate symbol $p \in \mathbf{Pred}$ with arity $n$, applied to terms $t_1,\dots,t_n$, which we write as $p(t_1,\dots,t_n)$. We will represent atoms by $H,A,B$;
    \item a query is a finite sequence of atoms, which we will represent by $Q,\bar{A},\bar{B}$;
    \item a clause is of the form $H\leftarrow \bar{B}$, where $H$ is an atom and $\bar{B}$ is a query;
    \item a program is a finite set of clauses, which we will represent by $P$.
\end{itemize}

\subsection{TSLD-resolution}

The operational semantics we will use is TSLD-resolution, first defined in \cite{BarbosaFloridoCosta22}. Here we present the definitions of a TSLD-step, a TSLD-derivation, and a TSLD-tree, all of which are necessary to understand how to detect type errors in programs and queries using this semantics.

\subsubsection{TSLD-derivation}

To compute, in logic programming, we need to have a program $P$ and a query $Q$. Then a TSLD-step, our basic step of computation, consists of selecting one atom $A$ in $Q$ and one clause $H \leftarrow \bar{B}$ in $P$, such that the predicate in $H$ is the same predicate as the one in $A$, unifying $H$ and $A$, and, if the result of the unification is a unifier $\theta$, applying $\theta$ to the remaining query together with $\bar{B}$. Note that unification is not defined for predicates, so we actually unify the terms in $H$ and $A$ pointwise, simultaneously.

\begin{definition}[TSLD-step]
Consider a non-empty query $Q = \bar{A_1},A,\bar{A_2}$ and a clause $c$ of the form $H \leftarrow \bar{B}$. Suppose that $A$ unifies (using typed unification) with $H$ and let $\theta$ be an mgu of $A$ and $H$. $A$ is called the {\em selected atom} of $Q$. Then we write $$\bar{A_1},A,\bar{A_2} \underset{c}\implies \theta(\bar{A_1},\bar{B},\bar{A_2})$$ and call it a {\em TSLD-step}. $H \leftarrow \bar{B}$ is called its {\em input clause}.
If typed unification of $A$ and $c$ outputs $wrong$ (or $\mathit{false}$) we write the TSLD-step as $Q \underset{c}\implies wrong$ (or $Q \underset{c}\implies false,\bar{A_1},\bar{A_2}$). 
\end{definition}

Note that when the result of unification is \textit{wrong} we are saying that there is a type error. Since this is what we are trying to detect, once we get a \textit{wrong} we delete the other atoms in that branch of the TSLD-tree. This behavior is different from \textit{false}, where we must continue with the other atoms in the query, because one of them may result in \textit{wrong}.

In this definition we assume that $A$ is variable disjoint with $H$. It is always possible to rename the variables in $H \leftarrow \bar{B}$ in order to achieve this, without loss of generality. Also, note that queries can be sequences of atoms and $\mathit{false}$. However, in a TSLD-derivation step, we must choose an atom of the query, so we cannot choose $\mathit{false}$.

\begin{definition}[TSLD-derivation]
Given a program $P$ and a query $Q$ a sequence of TSLD-steps from $Q$ with input clauses of $P$ reaching the empty query, $\mathit{false}$, or $wrong$, is called a {\em TSLD-derivation} of $Q$ in $P$.
\end{definition}

Since we can have several unifications resulting in $\mathit{false}$, in fact, we are considering that the comma is idempotent, so $\mathit{false},\mathit{false} \Leftrightarrow \mathit{false}$.
We will omit the program $P$ whenever it is clear from the context and we will omit the input clauses for TSLD-steps when they are irrelevant.
In \cite{BarbosaFloridoCosta22}, the authors prove that the selection of the atom and the mgu do not affect the result of a derivation. The selection of the clause from $P$ gives rise to the concept of TSLD-tree.

\subsubsection{TSLD-tree}

Since we can choose a clause from the program to unify with the selected atom of the query, we should consider all possible choices of said clause.
We say that a clause $H \leftarrow \bar{B}$ is {\em applicable} to an atom $A$ if $H$ and $A$ have the same predicate symbol with the same arity.
\begin{definition}[TSLD-tree]
Given a program $P$ and a query $Q$, a {\em TSLD-tree} for $(P,Q)$ is a tree where each branch is a TSLD-derivation of $Q$ in $P$, and the query in every node has a child for each clause from $P$ applicable to the selected atom of that query. 
\end{definition}

In order for the TSLD-tree to be unique (up to reordering of branches), we need to fix the order of selection of the atom from $Q$. Prolog uses the leftmost selection rule, where one always selects the leftmost atom in the query, and, since the selection rule does not change the result of a TSLD-derivation, we will use this selection rule in the rest of the paper.

The following two definitions are what allows us to detect type errors in logic programs using TSLD-trees.

\begin{definition}[Generic Query]
Let $Q$ be a query and $P$ a program. We say that $Q$ is a {\em generic query} of $P$ iff $Q$ is composed of a single atom of the form $p(X_1,\dots,X_n)$ for some predicate symbol $p$ that occurs in the head of at least one clause in $P$, and $X_1,\dots,X_n$ are all distinct variables.
\end{definition}

\begin{definition}[Blamed Clause]
Given a program $P$ and a query $Q$, a clause $c$ is a {\em blamed clause} of the TSLD-tree for $(P,Q)$ if all branches where $c$ is an input clause have leaf $wrong$.
\end{definition}
The blamed clause is a clause in the program which has a type error. A similar notion was first defined for functional programming languages with the blame calculus \cite{DBLP:conf/esop/WadlerF09}.

\begin{definition}[Type Error in the Program]
Suppose we have a program $P$. We say that $P$ has a {\em type error} if at least one clause \emph{c} is the blamed clause for the generic query $Q$ for the predicate in the head of \emph{c}.
\end{definition}

Note that if a program does not have a type error, then no generic query has a blamed clause. Intuitively, having a type error in the program means that somewhere in the program we will perform typed unification between two terms that cannot have the same type.

\begin{definition}[Ill-Typed Query]
Let $P$ be a program and $Q$ be a query. If there is no type error in $P$ and the TSLD-tree $(P,Q)$ is such that all branches have leaf \textit{wrong}, then we say that $Q$ is \textit{ill-typed} with respect to $P$.
\end{definition}

In \cite{BarbosaFloridoCosta22}, the authors defined a semantics for logic programs and queries, where the concepts of type error in a program and type error in a query become clear.
This semantics creates typed models for programs using the set obtained from the fixed point operator $T_P$, which is already used in the traditional semantics of logic programming. The typed interpretations generated from $T_P$ may model the program or not, and in fact a program with no such interpretation being a model has a type error.


\section{Maude}
\label{sec:maude}

Maude is a declarative language based on rewriting logic \cite{unified-tcs}, which contains order-sorted equational logic. 
Transition rules in rewriting logic allow for nondeterminism whereas equations and axioms allow for canonical algebras.
An equational program is a functional program in which a functional expression 
is deterministically evaluated until normal form modulo some commonly occurring axioms such as associativity and commutativity. 
Transition rules allow 
for a reachability graph between 
equivalence classes of terms.

Maude syntax is almost self-explanatory. 
Consider 
the following Maude functional module defining a meta representation for terms that
allows an arbitrary number of constant and function symbols, taken from \cite{DEEMMRTJLAMP20}.

\begin{verbatim}
fmod TERM is protecting NAT + QID .
   sort Var . op x{_} : Nat -> Var .
   sorts Term NvTerm . subsorts Qid < NvTerm < Term . subsort Var < Term .
   op _[_] : Qid NeTermList -> NvTerm .
   sort NeTermList . subsort Term < NeTermList .
   op _,_ : NeTermList NeTermList -> NeTermList [assoc] .
endfm
\end{verbatim}

\noindent
This module does not contain any equation, only the associativity structural axiom for the list concatenation operator \verb!_,_!. 
For example, 
a term \texttt{f(g(x, b, y), k(z))} is
here meta represented as the term \texttt{’f[’g[x\{1\},’b,x\{2\}],’k[x\{3\}]]} of sort \texttt{NvTerm} for non-variable terms, which is a subsort of the top sort \texttt{Term}. The term \texttt{’b} has sort
\texttt{Qid} imported from module \texttt{QID} and variables \texttt{x,y,z} are represented as
\texttt{x\{1\},x\{2\},x\{3\}}
of sort \texttt{Var}, a subsort of \texttt{Term}.

Equations are of the form
$\texttt{ceq}\ t = t' \ \texttt{if}\ u_1 = v_1 \wedge \ldots \wedge u_n = v_n$, or \texttt{eq} if the equation is unconditional,
and are applied from left to right modulo the structural axioms until normal form.
For example, the following functional module excerpt
from \cite{DEEMMRTJLAMP20}
defines Martelli-Montanari unification using the previous meta-representation module.

\begin{verbatim}
fmod LP-UNIFICATION is
   op unify : NeTermList NeTermList Substitution -> [Substitution] .
   eq unify(C, C, S) = S .
   eq unify(F[NeTL1], F[NeTL2], S) = unify(NeTL1, NeTL2, S) .
   eq unify(V, T2, (V -> T1) ; S) = unify(T1, T2, (V -> T1) ; S) .
   eq unify(NvT1, V, (V -> T2) ; S) = unify(NvT1, T2, (V -> T2) ; S).
   ceq unify(V, T, S) = (V -> T) ; S if not occur(V, S) .
   ceq unify(T, V, S) = (V -> T) ; S if not occur(V, S) .
   ceq unify((T1,NeTL1), (T2, NeTL2), S) = unify(NeTL1, NeTL2, S')
      if S' := unify(T1, T2, S) .
endfm\end{verbatim}

\noindent
Note that failure to unify means the resulting normal form of a call to symbol \texttt{unify} is not of sort \texttt{Substitution} but of its kind \texttt{[Substitution]}.
The condition $T_1 := T_2$ means that the normal form of $T_2$ must match the pattern $T_1$, which may contain extra variables not appearing in the lefthand side of the equation.

Transition rules are of the form 
$\texttt{crl}\ t \Rightarrow t'\ \texttt{if}\ u_1 = v_1 \wedge \ldots \wedge u_n = v_n$,
with \texttt{rl} if the rule is unconditional.
For example, the following rewrite theory
excerpt from \cite{DEEMMRTJLAMP20} defines a one-rule Prolog interpreter
using a configuration 
of the form
\verb!< N | PL $ S | Pr >!
where \texttt{N} is a variable counter, \texttt{PL} is the current query (predicate list), 
\texttt{S} is the accumulated substitution, 
and 
\texttt{Pr} is the logic program\footnote{
A clause is encoded as a term $P \ \texttt{:-}\  \mathit{PL}$ where $P$ is a term of sort \texttt{Predicate} and
$\mathit{PL}$ is a term of sort \texttt{PredicateList}.
A fact $p(\ldots)$ is encoded as a clause $p(\ldots)\ \texttt{:-}\  \texttt{nil}$.}.

\begin{verbatim}  
mod LP-SEMANTICS is
sort Configuration .
op <_|_$_|_> : Nat PredicateList Substitution Program -> Configuration .
crl < N1 | P1, PL1 $ S1 | Pr1 ; P2 :- PL2 ; Pr2 >
 => < N2 | PL3, PL1 $ S2 | Pr1 ; P2 :- PL2 ; Pr2 >
   if (P3 :- PL3) := rename(P2 :- PL2, N1)
   /\ S2 := unify(P1, P3, S1) /\ N2 := max(N1, last(P3 :- PL3)) .
endm
\end{verbatim}

\noindent
Note that this transition rule is 
nondeterministically exploring all those clauses \verb!P2 :- PL2! 
where \verb!P2! unifies with \verb!P1!.
This exploration relies on 
associativity of the operator \verb!;!.
Evaluating 
a query \texttt{PL}
using a logic program \texttt{Pr}
requires a command for reachability\footnote{Maude does not allow the underscore for unnamed variables as in logic programming but we use it in \texttt{search} commands in this paper for simplicity.}
\verb!search < PL | Pr > =>* < _ | nil $ S | _ >! 
where the configuration
\verb!< PL | Pr >!
will be properly expanded into an initial configuration
and
\texttt{S} would be the computed substitution.

\vspace{-11pt}
\section{\textsf{MaudeTypedLog} - A Typed Prolog Interpreter}
\vspace{-5pt}

\textsf{MaudeTypedLog}\footnote{Available at \url{https://github.com/egaltro/MaudeTypedLog}.} is a typed logic programming interpreter that 
extends
an interpreter
from 
\cite{DEEMMRTJLAMP20}
with
the typed unification algorithm 
of 
\cite{barbosa2024iclp}
and 
the TSLD-resolution 
of
\cite{BarbosaFloridoCosta22}.

\subsection{Implementation of the typed unification algorithm}\label{sec:unification}

We have implemented the typed unification algorithm of \cite{barbosa2024iclp} using  functional modules in Maude, since they are deterministic and return the unique normal form of a given initial term. Each inference rule of the unification algorithm  represents a step towards the final state of the unification and is encoded into Maude equations and specific data structures using associativity, commutativity, and identity.

For a unification of $t_1$ and $t_2$,
our implementation returns a pair of substitutions, one for terms and one for types, where each may be an error value.
The implementation is split in two parts: the first half corresponds to type unification, and the second half to term unification. We have an operator called \texttt{unify}, which receives as input some type equality constraints and some term equality constraints.

\begin{verbatim}
op unify : TermConstraintSet TypeConstraintSet -> SubstitutionPair .
eq unify(TCS, TYCS) = {getSubTe(unifyTe(TCS)) : getSubTy(unifyTy(TYCS))}.
\end{verbatim}

\noindent
This operator is defined by a single equation that uses two auxiliary operators \texttt{unifyTe} and \texttt{unifyTy} each of which corresponds to the unification of terms and types, respectively. 
Note that \texttt{unifyTy} corresponds to the rules 1-6 of 
Figure~\ref{ConsSolve} and \texttt{unifyTe} to rules 7-12
of 
Figure~\ref{ConsSolve}.

Each inference rule of the typed unification algorithm of
\cite{barbosa2024iclp}
could easily be encoded into one transition rule in Maude.
When no inference rule can be applied, we obtain the solution as a set of unification problems in solved form.
That is, unification problems are
of the form $X \mapsto t$, where
each variable has one and only one binding, and
we assume bindings are ordered 
in alphabetical order of the variable name.

However, we wanted a functional
encoding.
Since 
Maude does not follow 
any specific textual order 
to apply equations, 
in contrast to Haskell, 
 we encoded each inference rule
 into one conditional equation
 such that the conditions force only one equation to be applicable at a time. 
 Also, there is only one solved form modulo associativity, commutativity, and identity
 and we explicitly detect it, i.e., 
 the algorithm has finished and the current set of unification problems is the most general unifier.
A set of unification problems is in solved form by checking that a variable that occurs in the left-hand side of a constraint in the constraint set cannot occur anywhere else on the set. In order to verify this, we define the constraints as non-commutative and assume the variables are ordered. Then we add a rule for swapping the sides of a constraint according to the order on the variables. 

The possible outputs of \texttt{unifyTy} are either a type substitution or the error constant for type unification, \texttt{wrongType}. Similarly, the possible outputs of \texttt{unifyTe} are either a term substitution or the error constant for term unification, \texttt{falseSub}.
This differs from the failure of the unification algorithm of Section~\ref{sec:maude}.





We have adapted the meta representation of terms of Section~\ref{sec:maude} as follows.

\begin{verbatim}
subsorts QidC NatP < BaseTerm .
subsorts BaseTerm VarTerm FTerm ListTerm < Term < NeTermList < TermList .
\end{verbatim}

\noindent
We have \textit{base terms}, which are quoted identifiers \texttt{QidC} and natural numbers \texttt{NatP} in Peano notation, variable terms, function terms, and lists in the head-tail format. All of those are subsorts of the sort \texttt{Term}, which is a subsort of the non-empty term list \texttt{NeTermList}, which is itself a subsort of term list \texttt{TermList}. Floating-point numbers are not included, but they could be easily added to the interpreter.

In reality, our implementation 
is invoked with an initial set of term unification problems 
and we generate the initial set of type constraints. We define an operator \texttt{generate} that gets as input a term equality  and outputs type equality constraints as described by the rules in Figure \ref{ConsGen}.




\subsection{TSLD-resolution semantics}\label{sec:lp-semantics}

\begin{figure}[t]
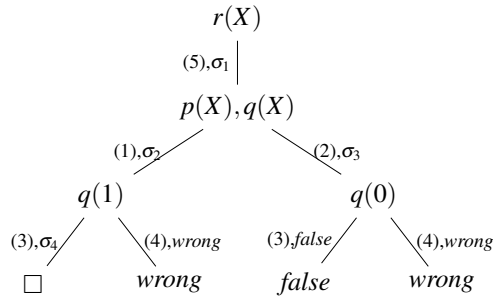


\texttt{crl [success] :}\\
        \texttt{< N1 | P1 PL1 \$ SB1 | PR1 ; (P2 :- PL2) ; PR2 - CL > =>}\\
        \texttt{< N2 | PL2 PL1 \$ {TS2? ; SB1} | PR1 ; (P2 :- PL2) ; PR2 - CL ; P2 :- PL2 >}\\
        \texttt{if P1 =/= falsePredicate /\symbol{92} P1 =/= wrongPredicate}\\
            \hspace*{.55cm}\texttt{/\symbol{92} \{TS2? : TYS2?\} := unify(P1, P2)}\\
            \hspace*{.55cm}\texttt{/\symbol{92} (TS2? =/= falseSub) /\symbol{92} (TYS2? =/= wrongType) .}\\[-.3cm]

\texttt{crl [false] :}\\
        \texttt{< N1 | P1 PL1 \$ SB1 | PR1 ; (P2 :- PL2) ; PR2 - CL > =>}\\
        \texttt{< N2 | PL1 falsePredicate \$ SB1}\\
        \hspace*{.84cm}\texttt{| PR1 ; (P2 :- PL2) ; PR2 - CL ; P2 :- PL2 >}\\
        \texttt{if P1 =/= wrongPredicate /\symbol{92} P1 =/= falsePredicate}\\
            \hspace*{.55cm}\texttt{/\symbol{92} \{TS2? : TYS2?\} := unify(P1, P2)}\\
            \hspace*{.55cm}\texttt{/\symbol{92} (TS2? == falseSub) /\symbol{92} (TYS2? =/= wrongType) .}\\[-.3cm]

\texttt{crl [wrong] :}\\
        \texttt{< N1 | P1 PL1 \$ SB1 | PR1 ; (P2 :- PL2) ; PR2 - CL > =>}\\
        \texttt{< N2 | wrongPredicate \$ SB1 | PR1 ; (P2 :- PL2) ; PR2 - CL ; P2 :- PL2 >}\\
        \texttt{if \{TS2? : TYS2?\} := unify(P1,P2) /\symbol{92} (TYS2? == wrongType) .}
\caption{Rules for TSLD-semantics}
\label{Rules}
\end{figure}

In order to implement an interpreter that follows the TSLD-resolution operational semantics for logic programs, we defined three rules in Figure~\ref{Rules} that execute TSLD-steps given a program and a query. These three rules make a case distinction over the possible \texttt{SubstitutionPair} that we get from unification. This is an extension of the Prolog interpreter defined in Section~\ref{sec:maude} that follows SLD-resolution as the semantics for Prolog. The main extension is the number of rules: in SLD-resolution we perform a unification that either returns a substitution or fails, but the failing case simply gets stuck, so there is a single rule. In TSLD-resolution we need to distinguish between $\mathit{false}$ and $wrong$, so the interpreter also distinguishes both cases. These are the two new rules for the semantics -- one for $\mathit{false}$ and one for $wrong$.

Every rule, as in the original interpreter of Section~\ref{sec:maude}, rewrites configurations to configurations until we get to a configuration that is somehow \textit{final}. However, configurations are slightly different, since they keep the list of selected clauses for each TSLD-step. 


\begin{verbatim}
op <_|_$_|_-_> : NatP PredicateList SubstitutionPair LPProgram LPProgram 
               -> LPConfiguration .
\end{verbatim}

\noindent
Note that each configuration has a natural number (to guarantee variable freshness), the current query, the current substitution (after \texttt{\$}), and the program and chosen clauses separated by \texttt{-}.

We have simplified the rules for presentation, in fact there is some renaming of variables and the new variable counter is calculated at each step as the largest variable index used in the rest of the configuration.


The unify algorithm ensures that each rule takes the first atom from the query and a clause from the program, such that the head of the clause has the same predicate as the selected atom, and unifies them. If the output from the unification algorithm returns \texttt{wrongType}, which corresponds to failing to unify the types, then we take a TSLD-step where unification of types failed as can be seen in rule \texttt{[wrong]}. It will return \texttt{wrongPredicate} in the resulting query and this TSLD-derivation is finished. If the output from unification returns \texttt{falseSub} but it does not contain \texttt{wrongType}, which corresponds to failing the unification of terms but succeeding for types, then we take a TSLD-step where unification of terms failed, adding the query \texttt{falsePredicate} to the end of the query and proceeding with this TSLD-derivation, as can be seen in rule \texttt{[false]}. If the output from unification is a pair of substitutions, which corresponds to a successful unification, then we add this substitution to the current one and continue the TSLD-derivation.

To show that the execution of these rules works as expected, we use Maude's \texttt{search} command. This command takes an initial term and a target pattern.
Our initial state is a configuration containing only a query and a program (which is expanded into a proper initial configuration) and in our target pattern the query is either the empty list of atoms, represent by \texttt{nil}, \texttt{wrongPredicate}, or \texttt{falsePredicate}. 
Maude applies all the rules in Figure \ref{Rules} non-deterministically from a \texttt{LPConfiguration} to another \texttt{LPConfiguration}. Therefore, we generate a TSLD-tree.

\begin{example} \label{ex1}
Consider the following logic program associated to identifier \texttt{ex1}.\\

\noindent
\texttt{p(1) :- nil ;;} \hfill (1)\\
\texttt{p(0) :- nil ;;} \hfill (2)\\
\texttt{q(1) :- nil ;;} \hfill (3)\\
\texttt{q(a) :- nil ;;} \hfill (4)\\
\texttt{r(X) :- p(X) , q(X) .} \hfill (5)\\

\noindent
Intuitively, this program is type error-free, since whenever we call the predicate \texttt{r}, it will search for terms accepted by both \texttt{p} and \texttt{q} and it finds one, specifically \texttt{1}. So the intuition is that if a clause \textit{can result} in something other than $wrong$, it does not have a type error.


The TSLD-tree for the generic query \verb!r(X)! is as follows,
where
we decorate each branch with the selected clause number and one of the following unifiers:
$\sigma_1=\{\}$,
$\sigma_2=\{X\mapsto 1\}$,
$\sigma_3=\{X\mapsto 0\}$,
and
$\sigma_4=\{\}$.

\begin{center}
\scalebox{.9}{
\begin{tikzpicture}
[level 1/.style = {sibling distance = 5cm},
level 2/.style = {sibling distance = 4cm},
level 3/.style = {sibling distance = 2cm}]
    \node {$r(X)$} [level distance = 1.3cm]
        child {node {$p(X),q(X)$}
            child {node {$q(1)$} 
                child {node {$\square$} edge from parent node [left,scale = .75] {(3),$\sigma_4$}
                }
                child {node {\textit{wrong}} edge from parent node [right,scale = .75] {(4),\textit{wrong}}
                }
            edge from parent node [left,scale = .75] {(1),$\sigma_2$}
            }
            child {node {$q(0)$}
                child {node {\textit{false}} edge from parent node [left,scale = .75] {(3),\textit{false}}
                }
                child {node {\textit{wrong}} edge from parent node [right,scale = .75] {(4),\textit{wrong}}
                }
            edge from parent node [right,scale = .75] {(2),$\sigma_3$}
            }
        edge from parent node [left,scale = .75] {(5),$\sigma_1$}
        };  
\end{tikzpicture}}
\end{center}


To compare \textsf{MaudeTypedLog} and the TSLD-resolution, we present the result for \texttt{search <r(X) | ex1> =>! <\_ | Out \$ \_ | \_ - Path >}:

\noindent
\begin{verbatim}
Solution 1:
Out -> nil
Path -> r(X) :- p(X),q(X) ; p(1) :-nil ; q(1) :-nil
Solution 2:
Out --> wrongPredicate
Path --> r(X) :- p(X),  q(X) ; p(1) :- nil ; q(a) :- nil
Solution 3:
Out --> falsePredicate
Path --> r(X) :- p(X),  q(X) ; p(0) :- nil ; q(1) :- nil
Solution 4:
Out --> wrongPredicate
Path --> r(X) :- p(X),  q(X) ; p(0) :- nil ; q(a) :- nil    
\end{verbatim}

It is clear that there is a solution for each leaf of the TSLD-tree and they correspond one-to-one, both in the result obtained and in the selected clauses.
\end{example}

\begin{example} \label{ex2}
Consider the following logic program associated to identifier \texttt{ex2}.\\

\noindent
\texttt{p(1) :- nil ;;}  \hfill (1)\\
\texttt{q(a) :- nil ;;}  \hfill (2)\\
\texttt{q(X) :- p(a) .} \hfill (3)\\

In order to formally type check, we need to create configurations with all possible generic queries, apply the rules until no more apply, and then gather all pairs of output and the selected clauses to get to that output. So, in short, each of these pairs is the result of one TSLD-derivation on the TSLD-tree. Then, a type error is identified if for some clause of the program, whenever it is a selected clause, the output is \texttt{wrongPredicate}, which is what happens for clause (3).
The TSLD-tree 
is
(we decorate each branch only with the clause number):

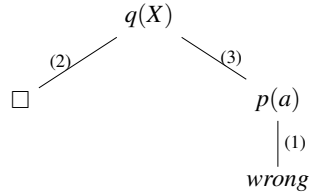
\begin{center}
\scalebox{.85}{
\begin{tikzpicture}
[level 1/.style = {sibling distance = 4cm},
level 2/.style = {sibling distance = 4cm},
level 3/.style = {sibling distance = 2cm}]
    \node {$q(X)$} [level distance = 1.3cm]
            child {node {$\square$} edge from parent node [left,scale=.75] {(2)}
            }
            child {node {$p(a)$} 
                child {node {$wrong$} edge from parent node [right,scale=.75] {(1)}
                }
            edge from parent node [right,scale=.75] {(3)}
            };
\end{tikzpicture}}
\end{center}

We can see that whenever clause (3) is selected for the generic query $q(X)$ we go to $wrong$, so (3) has a type error. \textsf{MaudeTypedLog} detects type errors in a program through an operator \texttt{hasTypeError} defined via equations, with signature \texttt{hasTypeError : LPProgram -> LPProgram}, that outputs the clauses that have a type error. We can see this in the solved form of the following 
\texttt{reduce} command, where \texttt{ex2} is the program above encoded in Maude:

\noindent
\begin{verbatim}
red hasTypeError(ex2) .
result Clause: q(X0) :- p(a)    
\end{verbatim}
\end{example}

\section{\textsf{MaudeTypedLog} Infrastructure} 

\textsf{MaudeTypedLog}
is implemented in Maude
and consists of
several Maude (functional and system) modules arranged in four 
levels.
The first level is the 
typed unification of Section~\ref{sec:unification} as a functional module.
The second level is the 
TLSD-resolution of Section~\ref{sec:lp-semantics} as a rewrite theory with transition rules.
The third level is the
functional module that performs calls
to the meta-level of Maude using both the first and second level,
specifically by invoking the \texttt{metaSearch} command for exploring the TSLD-tree as a data structure.
The fourth level 
is a command line interface (CLI) 
using object-oriented modules
that
allows users to interact with the tool  by means of simple, user-friendly commands, as illustrated below
by taking advantage
of the most recent I/O capabilities of Maude. 

Many applications developed in Maude
are written as 
object-oriented modules in which classes and subclasses are declared,
with the usual support for inheritance, dynamic binding, etc.  A class
is declared with syntax 
$
\mathtt{class}\;C \mid a_1 : S_1, \ldots, a_n : S_n
$
where $C$ is the name of the class,
$a_i$ are attribute identifiers, and $S_{i}$ are the sorts of the
corresponding attributes.  The objects of a class $C$ are then
record-like structures of the form 
$
< O : C \mid a_1 : v_1, \ldots, a_n : v_n > 
$
where $O$ is the identifier of the
object, and $v_i$ are terms of corresponding sorts $S_i$ that represent
the current values of its attributes.
Exchanged messages are declared using the \texttt{msg} keyword (similar to the \texttt{op} keyword)
and must include the addressee of a message as well as the sender's object identifier.

Each state of a concurrent object-oriented Maude system is a configuration consisting of a multiset of objects and
messages, built up with an empty syntax \texttt{\_\,\_} 
associative and commutative multiset union operator.  
Such configurations evolve by applying transition rules of the following two forms:

\begin{center}
\begin{tabbing}
\ \ \ \ \=\texttt{crl} \= 
\(< O : C \mid \mathit{atts} >  m(O,\vec{v})\)\\
\> \(\Rightarrow\) \> \(< O : C \mid\mathit{atts}' >\) 
\(< Q_1:C''_1\mid\mathit{atts}''_1 >\ldots < Q_p:C''_p\mid\mathit{atts}''_p >\) 
\(M'_1\ldots M'_q\)\\
\> \(\texttt{if }\mathit{Cond}\) .
\end{tabbing}
\end{center}
\begin{center}
\begin{tabbing}
\ \ \ \ \=\texttt{crl} \= 
\(< O : C \mid \mathit{atts} >\)\\
\> \(\Rightarrow\) \> \(< O : C \mid\mathit{atts}' >\) 
\(< Q_1:C''_1\mid\mathit{atts}''_1 >\ldots < Q_p:C''_p\mid\mathit{atts}''_p >\) 
\(M'_1\ldots M'_q\)\\
\> \(\texttt{if }\mathit{Cond}\) .
\end{tabbing}
\end{center}


\noindent
The first type of rules describes the concurrent transitions 
of an actor reacting to a message $m(O,\vec{v})$ addressed to him
where he can change the internal state of its attributes from 
$\mathit{atts}$ to $\mathit{atts}'$,  create new actors
$Q_{1}, \dots,Q_{p}$,
and send new messages 
$M'_1\ldots M'_q$ to other actors.  
The second type of
rule generalizes to actors 
that can change its internal attributes without the
prompting of a message. 

The object-oriented modules
also allow access to Maude external objects.
Transition rules interacting with external objects works
just like regular rewriting
but some messages are sent, or received, from such external objects. 
Those messages are processed  asynchronously by the external objects, so objects in Maude are not altered.
Some external objects in Maude are TCP/IP sockets, 
standard I/O streams, files and directories, UNIX processes, real (system) time, and meta-interpreters.
Configurations that want to communicate with external objects must
contain at least one \emph{portal}, given the following declarations.

\begin{verbatim}    
sort Portal . subsort Portal < Configuration . 
op <> : -> Portal [ctor portal] .
\end{verbatim}

In the MaudeTypedLog tool there are several programs defined that the user can load to play with.
The user can load the program with the command \texttt{load program <N>} and see the loaded program with the command \texttt{show program} as follows. This is the program in Example \ref{ex1}.

\begin{verbatim}
> load program 1
Loaded program 1
> show program
Program clauses:
p(1) :- .
p(0) :- .
q(1) :- .
q(a) :- .
r(X0) :- p(X0), q(X0).
\end{verbatim}

Once the program has been loaded, the user can check whether the program has any type errors, or not, by executing the command \texttt{check}.

\begin{verbatim}
> check
Type checking program...
Program is type safe
\end{verbatim}

When the user executes the command \texttt{check}, Maude executes the rule shown below and calls the function \texttt{printCheckMain} with the loaded program \textit{P}, which will print the result of the verification process.
The \texttt{hasTypeError} function obtains the generic queries for the predicates of the loaded program and runs all the queries, gathering the results.
Every query returns a pair \textit{(output, path)} in which the \textit{output} is the result of the TSLD-derivation, and the \textit{path} is an ordered list of the clauses used to reach to that output.

\begin{verbatim}
rl < O : VM | action : check, program : P >
=> < O : VM | action : idle, program : P >
   write(stdout, O, "Type checking program...\n"+ printCheckMain(P)) .
op printCheckMain : Program -> String .
eq printCheckMain(P) = printCheck(hasTypeError(P)) .
\end{verbatim}

There are two generic queries in Example~\ref{ex2}, \texttt{p(X0)} and \texttt{q(X0)}. We have to traverse the TSLD-tree for these generic queries and gather the outputs and selected clauses to get there.
Type checking this program detects a type error in clause (3) of Example~\ref{ex2}, as can be seen in the code excerpt below:

\begin{verbatim}
Loaded program 2
> check
Type checking program...
q(X0) :- p(a). Type error found
\end{verbatim}


The user can also create a query for the loaded program, just like Prolog.
To do so, the user only has to type the command \texttt{query (<QUERY>)} in MaudeTypedLog.
We have queried the term \texttt{(r(1))} in the Example \ref{ex1} and it should return \texttt{True} as there is a path that leads to unify the query with the program, although there are paths leading to a \texttt{false} or \texttt{wrong} output.

\begin{verbatim}
Loaded program 1
> query (r(1))
Executing query:
----------------
True
\end{verbatim}


\section{Conclusion and Future Work}

We have implemented a logic programming interpreter that follows the TSLD-resolution as the operational semantics for logic programs. The interpreter uses a typed unification algorithm as the basic computation unit, and allows for the detection of type errors both in programs and in queries.

We have tested this tool with some example programs, showing it is possible to detect type errors, and the results obtained are consistent with the theoretical ones.

There are still some features we wish to include in this interpreter. In the future, we plan to expand type checks to include built-in arithmetic predicates of logic programming, such as the predicate \texttt{is/2}. Also, we want to make it possible to customize the data types used for type verification, by allowing the programmer to add types for constructors, for instance binary trees.

We also want to apply this interpreter to a large set of predicates, including the libraries available in Prolog compilers like YAP and SWI-Prolog. The reason we have yet to test these libraries is the heavy use of built-in and arithmetic predicates, whose type information still needs to be correctly considered in order to effectively type check the program they contain.


\bibliographystyle{eptcs}
\bibliography{bibliography}
\end{document}